\documentclass[final,5p,times,twocolumn]{elsarticle}
\usepackage{algorithmic}
\usepackage{textcomp}
\usepackage{amsmath,amssymb,amsfonts}
\usepackage{todonotes}
\usepackage{cuted}
\usepackage{circuitikz}
\usepackage{color}
\usepackage{tikz}
\usepackage{hyperref}
\usepackage{mathtools}
\usepackage{wrapfig}
\usepackage{bm}
\usepackage{soul}
\usepackage{graphicx}
\usepackage{natbib}
\usepackage{threeparttable}

\journal{Journal of Materials Science: Materials in Electronics}
\begin{document}

\begin{frontmatter}

\title{Raman Strain-Shift Measurements and Prediction from First-Principles\\ in Highly-Strained Silicon}
\author[inst1]{Nicolas~Roisin}
\ead{nicolas.roisin@uclouvain.be}

\author[inst2]{Marie-Stéphane~Colla}
\author[inst1]{Jean-Pierre~Raskin}
\author[inst1]{Denis~Flandre}

\address[inst1]{Institute of Information and Communication Technologies, Electronics and Applied Mathematics (ICTEAM), UCLouvain, Place du levant 3, 1348 Louvain-la-Neuve, Belgium}
\address[inst2]{Institute of Mechanics, Materials and Civil Engineering (IMMC), UCLouvain, Place du levant 2, 1348 Louvain-la-Neuve, Belgium}
     

\begin{abstract}
This work presents how first-principles simulations validated through experimental measurements lead to a new accurate prediction of the expected Raman shift as a function of strain in silicon. Structural relaxation of a strained primitive cell is first performed to tackle the relative displacement of the silicon atoms for each strain level. Density Functional Perturbation Theory (DFPT) is then used to compute the energy of the optical phonon modes in highly-strained silicon and retrieve the strain-shift trend.
The simulations are validated by experimental characterization, using scanning electron microscopy (SEM) coupled with backscattering Raman spectroscopy, of silicon microbeams fabricated using a top-down approach. The beams are strained up to 2\% thanks to the internal tensile stress of silicon nitride actuators, allowing a validation of the perturbation theory in high-strain conditions.
The results are compared with the phonon deformation potentials (PDP) theory and the uncertainty caused by the various parameters found in the literature is discussed.
The simulated strain-shift coefficients of -175.77 cm$^{-1}$ (resp. -400.85 cm$^{-1}$) and the experimental one of -160.99 cm$^{-1}$ (resp. -414.97 cm$^{-1}$) are found for the longitudinal optical LO (resp. transverse optical TO$_1$) mode, showing good agreement.

\end{abstract}


\end{frontmatter}

\section{Introduction}
Raman spectroscopy is a well-known optical characterization technique used in material analysis. The light interactions with the chemical bonds or crystal lattice can be used to recover detailed information about the structure, the phase, the morphology or molecular interactions, for instance \cite{Loudon1963,Knight1989,Butler2016,Shvets2019}. It shows the advantage of fast and non-destructive measurements with micrometer spatial resolution \cite{Jones2019}. 
In a semiconductor crystal, the Raman spectrum is made of the combination of the different lattice vibration modes, i.e. phonons. As these modes depend on the crystal structure which is sensitive to strain and temperature, likewise the peaks of the Raman response \cite{Ganesan1970,Urena-Begara2018}. The characteristic of the crystal structure under strain can thus be used to estimate the strain level in a number of semiconductors such as silicon (Si) \cite{Ogura2006,Urena2013}, silicon carbide (SiC) \cite{Rohmfeld2002}, gallium nitride (GaN) \cite{Davydov1997}, germanium (Ge) \cite{Gassenq2017,Baranov2006}.

In silicon, the degeneracy of the three optical phonon peaks (around 520.7 cm$^{-1}$) is lifted under strain conditions. The intensity of the strain field can then be directly retrieved from the peak positions if the proper relations between the strain and the shifts of the peaks are known. However, the determination of this relation is often based on phonon deformation potentials (PDP) theory where the parameters are theoretically \cite{Nielsen1985} or experimentally deduced \cite{Chandrasekhar1978,Anastassakis1970,Anastassakis1990}. The different values of the parameters referenced in the literature induce an uncertainty on the strain-shift coefficients to be used. 
The density functional theory (DFT) and perturbation theory (DFPT) can be used to predict the variation in the phonon bands of a strained crystal \cite{Roisin2021,Baroni2001}. This \textit{ab initio} approach lifts the linear assumption often used in analytical models while reducing the uncertainty of the PDP parameters deduced experimentally.

In this work, a first-principles study is realized to compute the strain-shift relation of the optical phonon modes in a silicon crystal strained along the [110] direction. This study is then experimentally validated using scanning electron microscopy (SEM) and Raman measurements on highly-strained silicon microbeams fabricated using a top-down process.
The general theory of Raman spectroscopy is firstly recalled. The first-principles method to retrieve the phonons energy in strained silicon and its results are then explained. The experimental validation with the fabrication process and the characterization of the strained samples is lastly presented along with a discussion between the simulation and experimental results. 

\section{General theory of Raman spectroscopy}

The electric field at a position $\mathbf{r}$ associated with monochromatic light of frequency $\omega_i$ and amplitude $\mathbf{E}_0$ propagating in a direction $\mathbf{k}_i$ can be mathematically described as
\begin{equation}
    \mathbf{E}(\mathbf{k}_i,\omega_i)=\mathbf{E}_0 ~e^{i(\mathbf{k}_i\cdot \mathbf{r}-\omega_i t)}.
\end{equation}

Once the light interacts with the material, the electric moment $\mathbf{P}$ resulting from this interaction is given by \cite{Boyd2003}
\begin{equation}
    \mathbf{P}=\varepsilon_0~\chi~\mathbf{E}(\mathbf{k}_i,\omega_i),
\end{equation}
where $\chi$ is the susceptibility tensor, describing the response of the crystal to an electric field. 

In what follows, the theory is from \cite{DeWolf1996}. As the positions of the atoms affect the susceptibility, a Taylor series with regards to the vibration modes can be used:
\begin{equation}
    \chi=\chi_0+\left(\frac{\partial\chi}{\partial\mathbf{Q}_j}\right)\mathbf{Q}_j+ \left(\frac{\partial^2\chi}{\partial \mathbf{Q}_j\partial \mathbf{Q}_k}\right)\mathbf{Q}_j\mathbf{Q}_k+...,
\end{equation}
\begin{equation}
    \mathbf{Q}_j(\mathbf{q}_j,\omega_j)=\mathbf{Q}_{j,0} ~e^{\pm i(\mathbf{q}_j\cdot \mathbf{r}-\omega_j t)},
\end{equation}
where $\mathbf{Q}_{j}$ is the normal coordinate of the vibration mode characterized by a wavevector $\mathbf{q}_j$ and a frequency $\omega_j$.

Under the assumption of one-phonon interaction, the electric moment can be written as
\begin{equation}
    \mathbf{P}=\varepsilon_0~\chi_0~\mathbf{E}_0~e^{j(\mathbf{k}_i\cdot \mathbf{r}-\omega_i t)}+\left(\frac{\partial\chi}{\partial\mathbf{Q}_j}\right)~\mathbf{E}_0~\mathbf{Q}_{j,0}~e^{-i(\omega_i\pm\omega_j)t}e^{i(\mathbf{k}_i\pm\mathbf{q}_j)\cdot \mathbf{r}}.
\end{equation}
This expression highlights the three light components radiated at distinct frequencies when a monochromatic light interacts with the material.
The first term corresponds to the elastic scattering of the incident light known as Rayleigh scattering ($\omega_i$ frequency). The inelastic scattering with one absorbed phonon corresponds to the Stokes ($\omega_i+\omega_j$ frequency) or anti-Stokes scattering ($\omega_i-\omega_j$ frequency) if the phonon is emitted.

\begin{figure}[h!]
    \centering
    \includegraphics[width=1\linewidth]{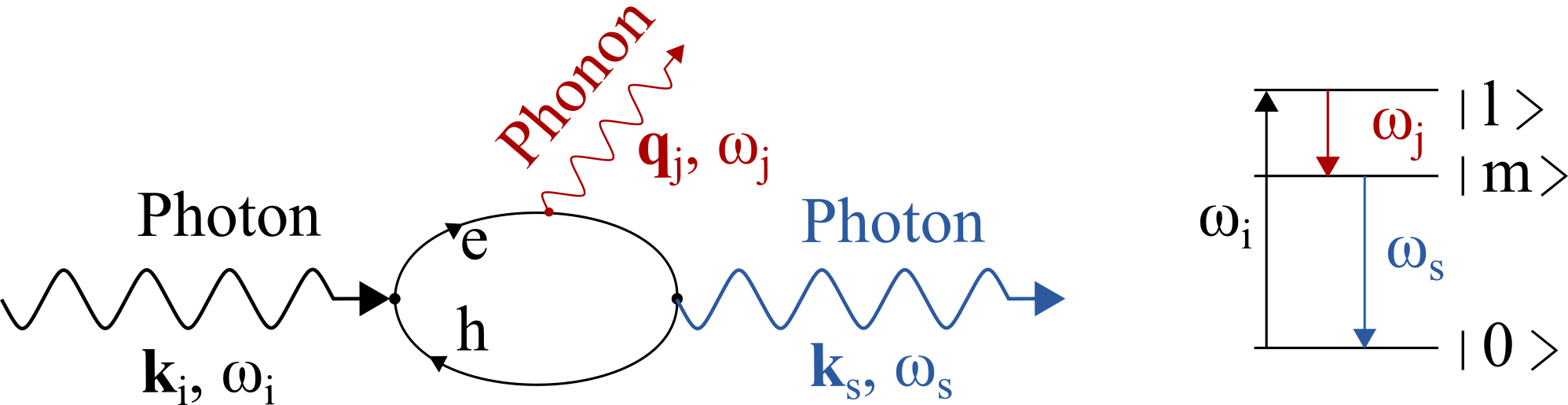}\vspace{0.5cm}
    \includegraphics[width=0.5\linewidth]{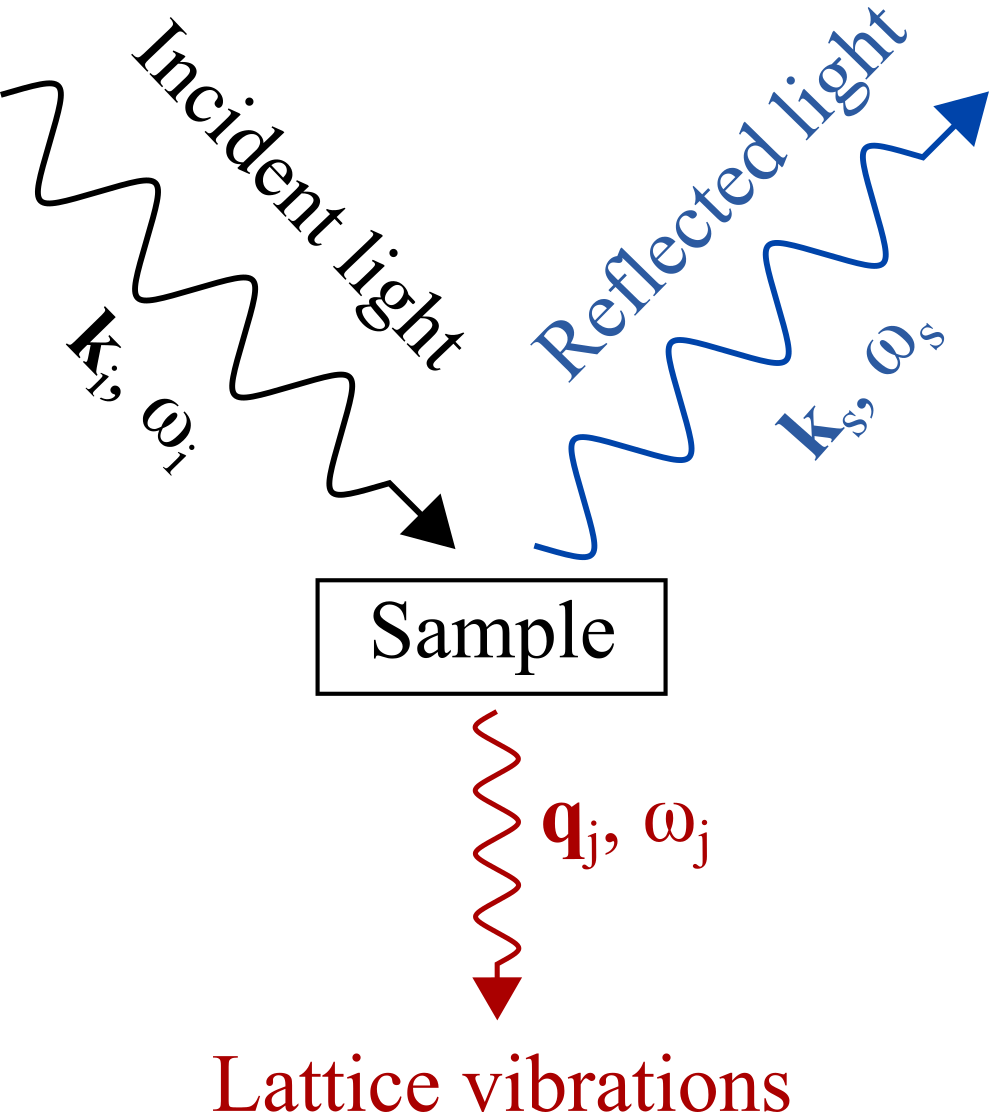}
    \caption{Feynmann diagram of the first-order Stokes inelastic scattering process. In this process, an absorbed incident photon ($\mathbf{k}_i,\mathbf{\omega}_i$) interact with the material that emits a phonon ($\mathbf{q}_j,\mathbf{\omega}_j$) and scattered photon ($\mathbf{k}_s,\mathbf{\omega}_s$). }
    \label{fig:feynmann}
\end{figure}

Fig. \ref{fig:feynmann} represents the Feynmann diagram for the Stokes scattering process. The incident photon ($\mathbf{k}_i,\mathbf{\omega}_i$) creates an electron-hole pair in the material due to the transition of an electron from state $\left|~0~\right\rangle$ to a state $\left|~l~\right\rangle$. 
The interaction between this electron and a phonon ($\mathbf{q}_j,\mathbf{\omega}_j$) moves the particle from state $\left|~l~\right\rangle$ to state $\left|~m~\right\rangle$. In this step, a phonon can be absorbed (anti-Stokes scattering with energy increase) or emitted (Stokes scattering with energy reduction).
Finally, the recombination of the electron-hole pair emits a photon ($\mathbf{k}_s,\mathbf{\omega}_s$) of frequency $\omega_s=\omega_i\pm\omega_j$.

In a crystal, the intensity of the Raman signal depends on the polarization vectors of the incident $\mathbf{e}_i$ and scattered $\mathbf{e}_s$ lights:
\begin{equation}
    I\propto \sum_j~|~\mathbf{e}_i~\cdot~\mathbf{R}_j~\cdot~\mathbf{e}_s~|,
\end{equation}
where $\mathbf{R}_j$ is the Raman tensor of the phonon mode $j$. Those tensors can be theoretically obtained based on the crystallography of the material \cite{Loudon1964}. For the diamond cubic structure of silicon in the crystal coordinates x=[100], y=[010] and z=[001], there are three tensors given by

\begin{equation}
\begin{array}{c}
       \mathbf{R}_x=\left(\begin{array}{ccc}
         0&0&0  \\
         0&0&d  \\
         0&d&0  \\
    \end{array}\right),\quad
        \mathbf{R}_y=\left(\begin{array}{ccc}
         0&0&d  \\
         0&0&0  \\
         d&0&0  \\
    \end{array}\right),\\
         \mathrm{and}\quad
        \mathbf{R}_z=\left(\begin{array}{ccc}
         0&d&0  \\
         d&0&0  \\
         0&0&0  \\
    \end{array}\right).
\end{array}
\end{equation}

In a Raman backscattering process from a (001) surface, the two transverse optical phonon modes (TO$_1$ and TO$_2$) polarized along $x$ and $y$ are given by $\mathbf{R}_x$ and $\mathbf{R}_y$, respectively. The tensor $\mathbf{R}_z$ corresponds to the longitudinal optical mode (LO) polarized along $z$. Basic knowledge about the polarization state of the excitation and scattered light is then needed to correctly observe the vibration modes of the crystal.

\section{Computation of the phonon energy in strained silicon}

In a silicon beam strained in the [110] direction, the strain tensor can be expressed as
\begin{equation}
    \textnormal{\textbf{E}}_{[110]} = \frac{1}{2} \left( 
    \begin{array}{ccc}
        (\varepsilon_{\perp}+\varepsilon_{\parallel}) & (\varepsilon_{\perp}-\varepsilon_{\parallel}) & 0 \\
        (\varepsilon_{\perp}-\varepsilon_{\parallel}) & (\varepsilon_{\perp}+\varepsilon_{\parallel}) & 0 \\
        0 & 0 & 2\varepsilon_{\parallel}
    \end{array}\right),
\end{equation}  
where $\varepsilon_{\perp}$ is the uniaxial strain applied on the material in the [110] direction while $\varepsilon_{\parallel}$ is the induced perpendicular strain. 
For elastic deformations, $\varepsilon_{\parallel}$ is proportional to $\varepsilon_{\perp}$:
\begin{equation}
    \varepsilon_{\parallel} = -\nu_{[110]}~ \varepsilon_{\perp},
\end{equation}
with $\nu_{[110]}$ the Poisson's ratio in the [110] direction:
\begin{equation}
    \nu_{[110]} = \frac{4C_{12} C_{44}}{2C_{11} C_{44} + (C_{11} + 2C_{12}) (C_{11} - C_{12})},
\end{equation}
where $C_{11}=165.77$~GPa, $C_{12}=63.93$~GPa, and $C_{44}=79.62$~GPa are the elastic constants of silicon \cite{Madelung1991}.

The strain tensor describes the macroscopic displacement of the crystal but also the lattice changes needed for the first-principles computation. The primitive vectors of the strained cell ($\mathbf{a}',~ \mathbf{b}'$and $\mathbf{c}'$) can then be retrieved as
\begin{equation}
   (\mathbf{a}'~ \mathbf{b}'~ \mathbf{c}')= \textnormal{\textbf{E}}_{[110]}\cdot (\mathbf{a}~ \mathbf{b}~ \mathbf{c}),
\end{equation}
where $\mathbf{a},~ \mathbf{b}$ and $\mathbf{c}$ are the vectors of the relaxed primitive cell given by
\begin{equation*}
    \mathbf{a}=\frac{a_0}{2}\left(\begin{array}{c}
         0  \\
         1 \\
         1
    \end{array}\right),~
        \mathbf{b}=\frac{a_0}{2}\left(\begin{array}{c}
         1  \\
         0 \\
         1
    \end{array}\right),~
        \mathbf{c}=\frac{a_0}{2}\left(\begin{array}{c}
         1  \\
         1 \\
         0
    \end{array}\right),
\end{equation*}
where $a_0=0.543$ nm is the lattice parameter of silicon \cite{Yasumasa1984}.

\subsection{Phonons energy computation}
Under the harmonic approximation, the frequencies of the phonon modes $\omega_{m\mathbf{q}}$ associated with the displacement $\mathrm{u}_\alpha(l\kappa)$ in direction $\alpha$ of the atom $\kappa$ with a mass $M_{\kappa}$ in unit cell $l$ and are obtained by solving the following eigenvalue problem:
\begin{equation}
    \sum_{\kappa'\beta}D_{\alpha\beta}(\kappa\kappa';\mathbf{q})~\gamma_{m\mathbf{q}}(\kappa'\beta)=\omega_{m\mathbf{q}}^2~\gamma_{m\mathbf{q}}(\kappa\alpha),
\end{equation}
where $\gamma_{m\mathbf{q}}(\kappa\alpha)$ are the phonon eigenvectors \cite{Born1954}.

The dynamical matrix $D_{\alpha\beta}(\kappa\kappa';\mathbf{q})$ is given by the second derivative of the potential energy $\Phi$ with respect to the displacement of the atoms $\kappa$ in unit cell $l$ and $\kappa'$ in unit cell $l'$:
\begin{equation}
D_{\alpha\beta}(\kappa\kappa';\mathbf{q}) = \frac{1}{\sqrt{M_\kappa M_{\kappa'}}}\sum_l\frac{\partial^2\Phi}{\partial \bm{\mathrm{u}}_\alpha(l\kappa)~\partial \bm{\mathrm{u}}_\beta(l'\kappa')}e^{i\mathbf{q}\cdot(\mathbf{x}(l\kappa)-\mathbf{x}(l'\kappa'))}.
\end{equation}

For the two-atoms unit cell of silicon, the number of eigenmodes is equal to six corresponding to the three acoustic and three optical modes of the crystal. However, the light excitation of the material mainly interacts with the optical modes with small wavevectors ($\mathbf{q}_j\rightarrow 0$) where the center of mass is stationary and the sub-lattices vibrate rigidly against each other \cite{Russell1965}.

In this work, the Density Functional Perturbation Theory (DFPT) as implemented in ABINIT \cite{Gonze2020,Romero2020} is used to compute the eigenvalues corresponding to the energy of the optical modes in strained silicon.

In practice, norm-conserving pseudo-potentials coming from PSEUDO DOJO are used \cite{Vansetten2018}, in the generalized-gradient approximation (GGA) from Perdew–Burke–Ernzerhof (PBE) with a plane-wave cutoff of 20 Ha \cite{Perdew1996}. An internal structural relaxation for the two atoms of the strained (fixed) unit cell is first performed for each strain level. An 8 x 8 x 8 Monkhorst–Pack k-point grid is then chosen to perform the self-consistent calculation of the total energy used as the starting point for the perturbation calculation at the $\Gamma$-point of the Brillouin zone ($\mathbf{q}=0$) \cite{Monkhorst1976}.

\begin{figure}[h!]
    \centering
    \includegraphics[width=1\linewidth]{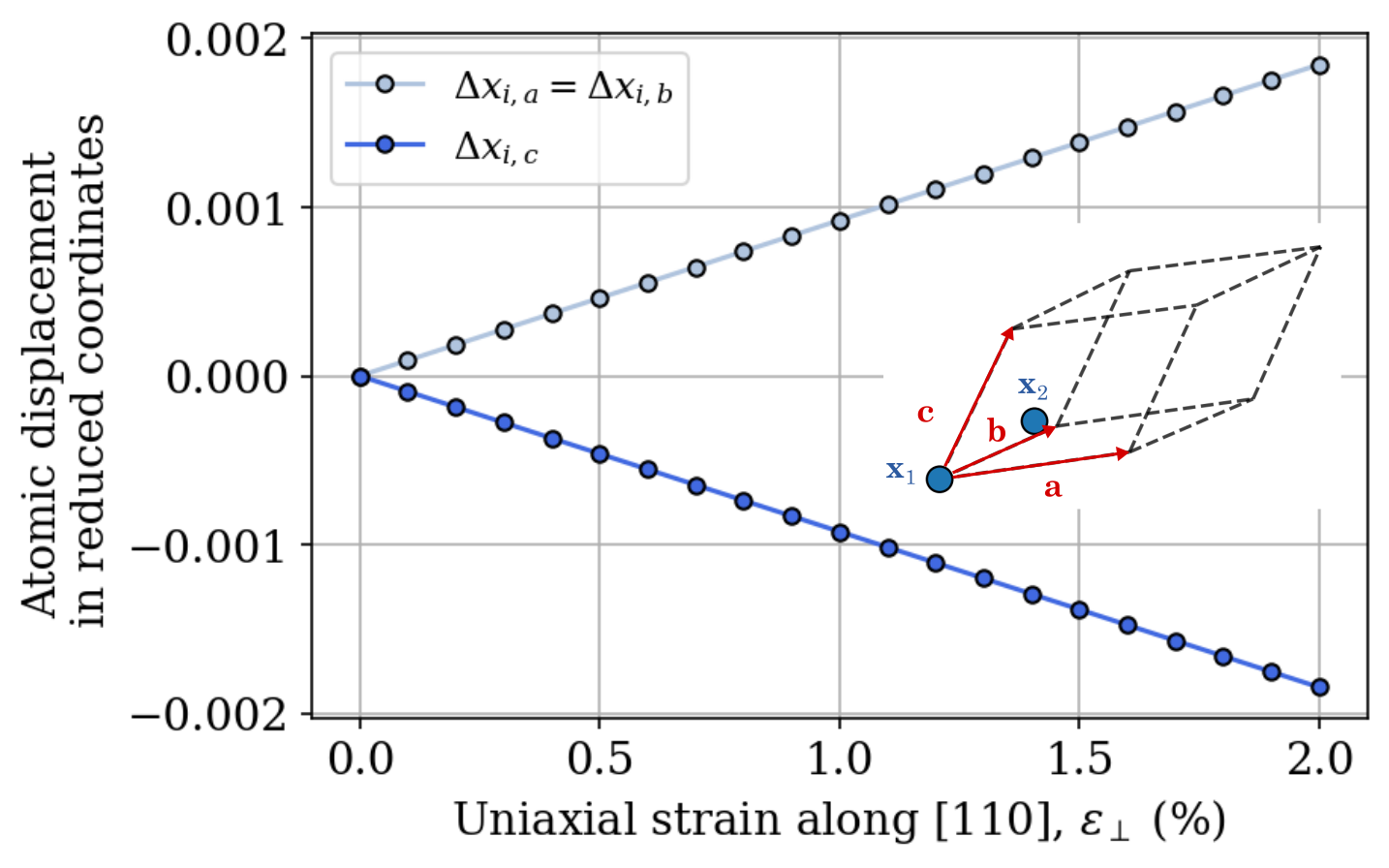}
    \caption{Results of displacement (in reduced coordinates) of the two silicon atoms inside the unit cell under strain conditions. The unit cell of silicon is represented next to the curves along with the three primitive lattice vectors (in red).}
    \label{fig:Displacements}
\end{figure}

In the internal relaxation step, the lattice vectors of the strained structures are calculated thanks to the strain tensor $\mathbf{E}_{[110]}$. However, the positions of the two silicon atoms need to be computed again to find the new equilibrium that minimizes the total energy of the strained system. The results of this step are presented in Fig. \ref{fig:Displacements}. Both atoms present the same displacement with regard to their relaxed positions. The relative displacements of the atoms inside the cell are negative for the coordinate along the primitive vector $\mathbf{c}$ aligned with the applied strain, i.e. the [110] direction, while the two other components along $\mathbf{a}$ and $\mathbf{b}$ show the opposite positive variation.

\begin{figure}[h!]
    \centering
    \includegraphics[width=1\linewidth]{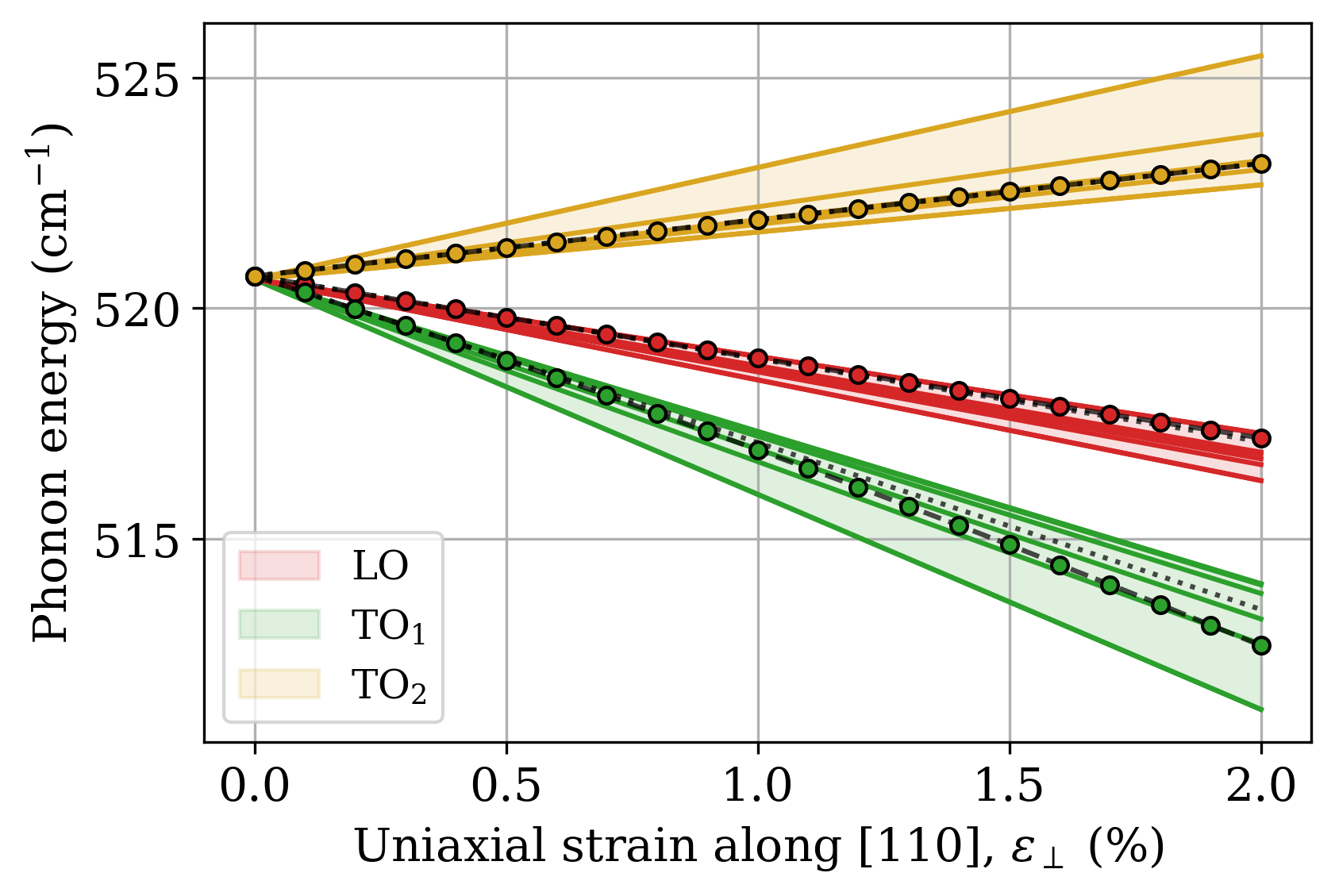}
    \caption{Results of the density functional perturbation simulations of the phonons energies for the three optical modes (LO, TO$_1$ and TO$_2$). The results (dots with dashed line) are compared with the PDP theory with several parameters given in Table \ref{tab:pdpcoeff} (continuous curves). A linear fit of the simulations is displayed (dotted line). The uncertainty caused by the different deformation potentials is highlighted with the filled area between the curves.}
    \label{fig:pdp}
\end{figure}

Once the strained equilibrium of the crystal has been retrieved, the phonon energy can be computed with DFPT computation at the center of the Brillouin zone ($\mathbf{q_j}=0$) for the three optical modes. The results of the simulations are shown in Fig. \ref{fig:pdp} and linear strain-shift coefficients of -175.77 cm$^{-1}$, -400.85 cm$^{-1}$ and 121.91 cm$^{-1}$ are found for the LO, TO$_1$ and TO$_2$ modes, respectively. The linear approximation is found to be acceptable for the LO and TO$_2$ modes while an error up to 0.078 cm$^{-1}$ at 2\% strain is made with the linear assumption for the TO$_1$ mode.

\subsection{Phonon Deformation Potential (PDP) theory}
Ganesan \textit{et al.} provide a theoretical model to predict the strain-shift of the frequencies of the triply degenerated optical modes in crystals with a diamond structure \cite{Ganesan1970}. 

In this phonon deformation potential (PDP) theory, the first-order perturbation of the dynamical matrix at $\mathbf{q}=0$ is computed and it has been shown that only three parameters ($p$, $q$ and $r$) are needed to compute the strain dependence:

\begin{equation}
    D_{\alpha\beta}=D_{\alpha\beta}^{(0)} +\sum_{\mu\nu}\frac{\partial D_{\alpha\beta}}{\partial\varepsilon_{\mu\nu}}\varepsilon_{\mu\nu}+...,
\end{equation}
where $\varepsilon_{\mu\nu}$ are the components of the strain tensor and $D_{\alpha\beta}(\kappa\kappa';0)$ is written as $D_{\alpha\beta}$.

For silicon, the first derivatives are then given by
\begin{equation}
    \frac{\partial D_{\mu\mu}}{\partial\varepsilon_{\mu\mu}}=p,\quad \frac{\partial D_{\mu\mu}}{\partial\varepsilon_{\nu\nu}}=q,\quad \frac{\partial D_{\mu\nu}}{\partial\varepsilon_{\mu\nu}}=r,
\end{equation}
while the other derivatives are zero.

The frequency shifts can then be computed by solving the so-called secular equation:

\begin{equation}
    \resizebox{0.45\textwidth}{!}{
    $\left|\begin{array}{ccc}
         p\varepsilon_{11}+q(\varepsilon_{22}+\varepsilon_{33})-\lambda& 2r\varepsilon_{12} &2r\varepsilon_{13} \\
         2r\varepsilon_{21}&p\varepsilon_{22}+q(\varepsilon_{11}+\varepsilon_{33})-\lambda  &2r\varepsilon_{23} \\
         2r\varepsilon_{31}&2r\varepsilon_{32}  &p\varepsilon_{33}+q(\varepsilon_{11}+\varepsilon_{22})-\lambda \\
    \end{array}\right|=0$},
\end{equation}
with $\lambda_i=\omega_i^2-\omega_0^2$.

Table \ref{tab:pdpcoeff} and Fig. \ref{fig:pdp} respectively summarize and represent the sets of PDP parameters found in the literature. The parameters for this work are computed using least-square fit of the DFPT result shown in Fig. \ref{fig:pdp}.  The colored areas in the figure illustrate the uncertainty induced by the choice of parameters. Indeed, the strain-shift coefficient for the longitudinal mode varies from -167 cm$^{-1}$ to -218 cm$^{-1}$ while the coefficient for the TO$_1$ (resp. TO$_2$) mode is between -330 cm$^{-1}$ (resp. 102 cm$^{-1}$ ) and -466 cm$^{-1}$ (resp. 242 cm$^{-1}$).
The uncertainty is found to be lower for the longitudinal mode than the transverse one as most of the experimental works are conducted on (100) silicon wafers with backscattered signals that usually present a strong LO peak.

\begin{table}[h!]
    \centering
    \begin{tabular}{c|ccc}
         Reference&$p/\omega_0^2$&$q/\omega_0^2$&$r/\omega_0^2$ \\\hline
         Anastassakis \textit{et al.} (exp.) \cite{Anastassakis1970}& -1.25&-1.87&-0.66\\
         Chandrasekhar \textit{et al.} (exp.) \cite{Chandrasekhar1978}&-1.49&-1.97&-0.61\\
         Nielsen \textit{et al.} (exp.) \cite{Nielsen1985}&-1.63 &-1.89 &-0.6\\
         Anastassakis \textit{et al.} (exp.) \cite{Anastassakis1990}&-1.85&-2.31&-0.71\\
         Kosemura \textit{et al.} (sim.) \cite{Kosemura2016}& -1.75 & -2.16 & /\\
        This work (sim.) &-1.67&-2.03&-0.67\\
    \end{tabular}

    \caption{Silicon phonon deformation potentials reported in the literature and obtained experimentally (exp.) or theoretically from first-principles simulations (sim.). The potentials referenced in this work are obtained by least-square fitting of the results shown in Fig. \ref{fig:pdp}.}
    \label{tab:pdpcoeff}
\end{table}

One of the limitations of this theory comes from the assumption that the shifts are linearly affected by the strain. Our simulation results present a slightly non-linear behavior compared with the PDP theory that assumed a linear response but the second-order term is found to be small compared with the first-order term in the strain range experimentally obtained (up to 2\%).

\section{Experimental validation}

\subsection{Fabrication of strained silicon microbeams}

The silicon microbeams are deformed without needing external actuation thanks to an on-chip tensile testing technique \cite{Coulombier2012,Escobedo-Cousin2011,KumarBhaskar2013}. The concept of this technique is to benefit from the high internal stress present in an \textit{actuator} layer made of silicon nitride in order to pull the \textit{specimen} layer, in silicon in this study \cite{Passi2012,Gravier2009}. 

\begin{figure}[h!]
    \centering
    \includegraphics[width=1\linewidth]{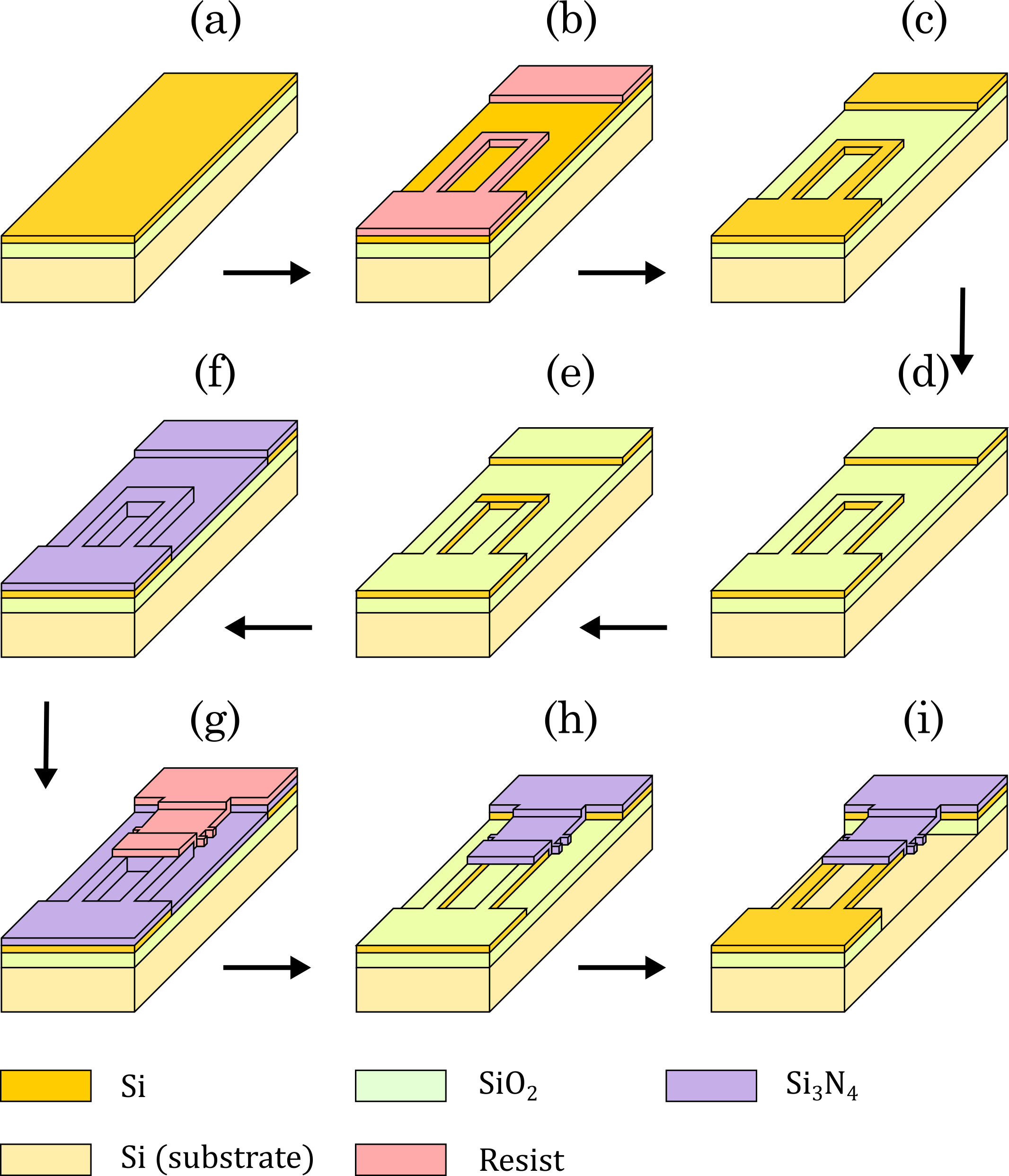}
    \caption{Overview of the fabrication process for the strained silicon microbeams. The complete description of the fabrication of the samples can be found in \cite{Bhaskar2011}.}
    \label{fig:process}
\end{figure}

The design shown in Fig. \ref{fig:process} has already been detailed in \cite{Bhaskar2011,Urena2012}. The on-chip structures are produced using microfabrication techniques as developed for microelectronics. The fabrication process starts with a p-type Si-on-Insulator (SOI) substrate with a top Si layer of 100 nm and a 1 µm-thick buried oxide (BOX) that will serve as a sacrificial layer. First, the top Si layer is patterned thanks to a photolithography step followed by a chlorine plasma etching to provide the shape to the microbeam specimens (Figs. \ref{fig:process} (a) to (c)). Then, a 20 nm-thick thermal oxide layer is grown on the patterned top-Si and locally etched away after photolithography and BHF solution in order to open windows at the overlapping area where the actuator will be in contact with the top Si microbeams (Figs. \ref{fig:process} (d) to (e)). Finally, the actuator layer made of low pressure chemical vapour deposited (LPCVD) silicon nitride film is deposited at 800°C on top of the partly oxidized top-Si and patterned using SF$_6$ dry etching (Figs. \ref{fig:process} (f) to (h)). The 20 nm-thick thermal oxide grown on the top Si serves as an etch stop layer during this processing step as the selectivity between Si$_3$N$_4$ and Si is not high enough under SF$_6$ plasma. The last step is the release of the global structure using hydrofluoric acid concentrated at 73\% in order to remove the sacrificial buried oxide lying under the specimen and actuator beams and in order to let the force equilibrium to operate (Figs. \ref{fig:process} (i)). After the release, the Si$_3$N$_4$ beam is pulling on the Si microbeam that is thus under tension. Typical dimensions are as follows: actuators are tens of µm-wide actuators, specimens are a few µm-wide and the length of both parts varies between tens to hundreds of µm.

\begin{figure}[h!]
    \centering
    \includegraphics[width=1\linewidth]{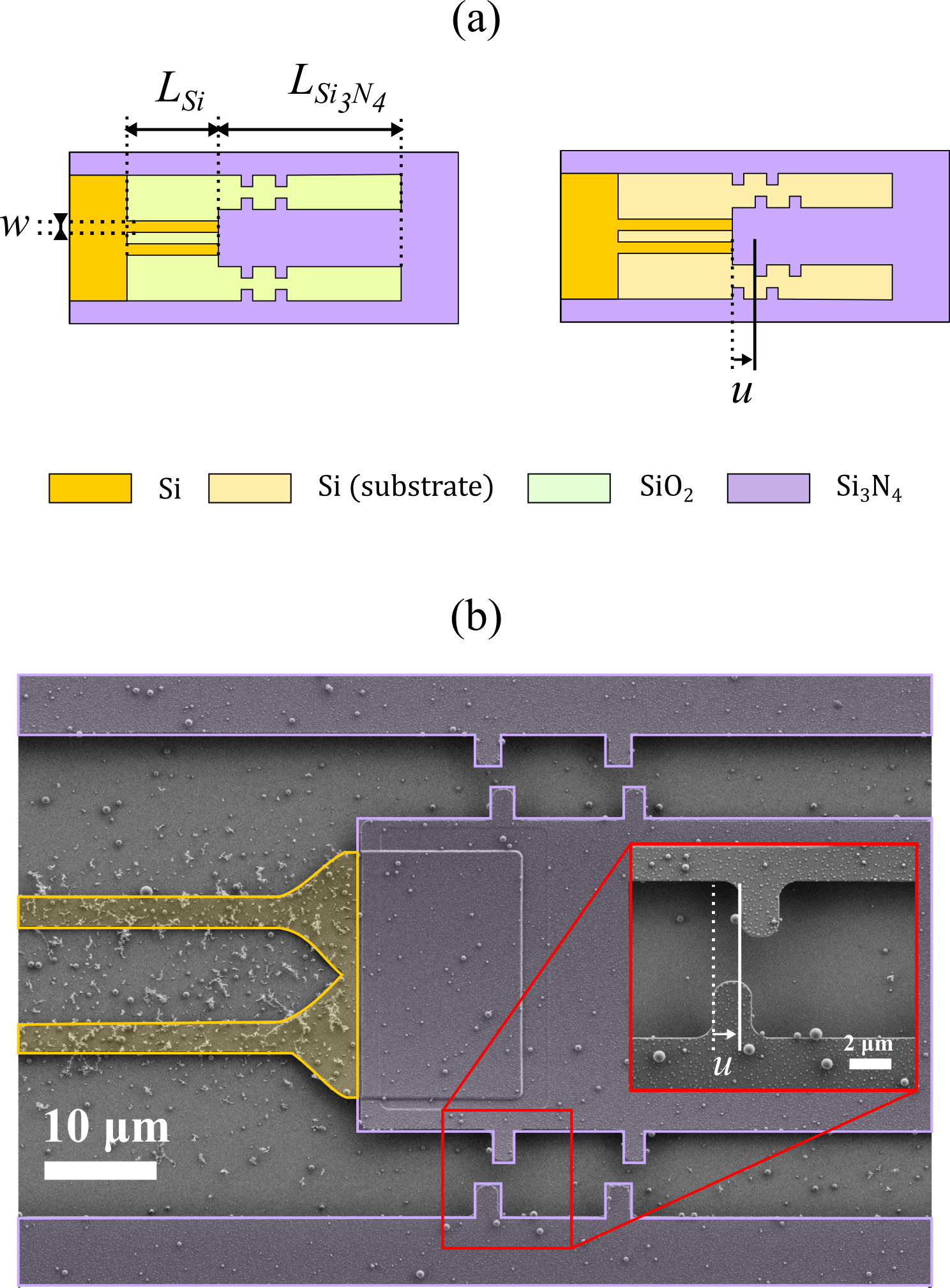}
    \caption{Illustration of (a) the diagram of the last step of the process (release) that builds the strain in the beams and, (b) the scanning electron microscope image (SEM) used to visually compute the strain based on the displacement of the cursors.}
    \label{fig:SEM_cursors}
\end{figure}

The displacement imposed to the Si specimen is measured inside a scanning electron microscope (SEM) thanks to cursors located along the beams, see Fig. \ref{fig:SEM_cursors}. As the top-Si layer does not contain any internal stress here, the mechanical strain $\varepsilon_{\perp}$ inside one Si microbeam can be obtained by:

\begin{equation} \label{eq:strain}
    \varepsilon_{\perp}=ln\left( \frac{L_{Si}+u}{L_{Si}} \right),
\end{equation}

where $L_{Si}$ is the initial length of the Si specimen and $u$ the displacement, see Fig. \ref{fig:SEM_cursors}.

\begin{figure}[h!]
    \centering
    \includegraphics[width=1\linewidth]{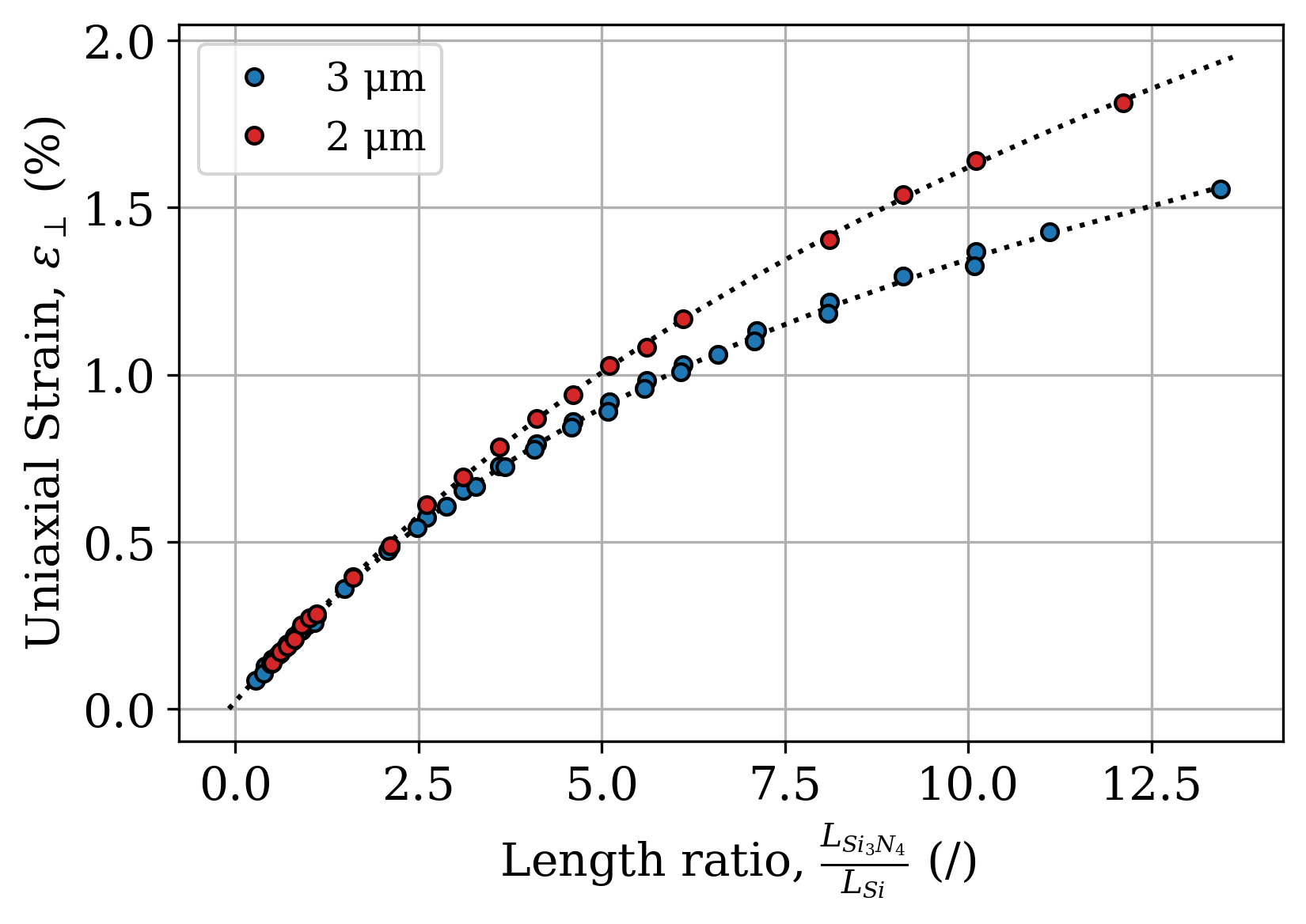}
    \caption{Uniaxial strain calculated in 2- and 3-$\mathrm{\mu}$m wide silicon microbeams using Eq. (\ref{eq:strain}) from SEM measurements of the displacement of the 150- and 200-$\mathrm{\mu}$m long silicon beams, when actuated by Si$_3$N$_4$ actuators of length varying from 55 $\mathrm{\mu}$m to 2000 $\mathrm{\mu}$m.} 
    
    \label{fig:SEM_results}
\end{figure}

On a 2 cm$^2$ die, several thousands of such structures are present with different sets of dimensions in order to apply varying strain levels in each Si microbeam. Using Eq. (\ref{eq:strain}), we obtain the graph presented in Fig. \ref{fig:SEM_results} where one point represents the strain in one Si microbeam. In this work, 2 and 3-$\mathrm{\mu}$m wide Si specimens with lengths $L_{Si}$ varying between 150 and 200 $\mathrm{\mu}$m have been characterized. The actuator length $L_{Si_3N_4}$ varies from 55 µm to 2000 µm to tune the length ratio that will define the strain in the Si microbeams. A maximum strain of about 2\% is reached here.

\subsection{Raman measurements}

\begin{figure}[h!]
    \centering
    \includegraphics[width=1\linewidth]{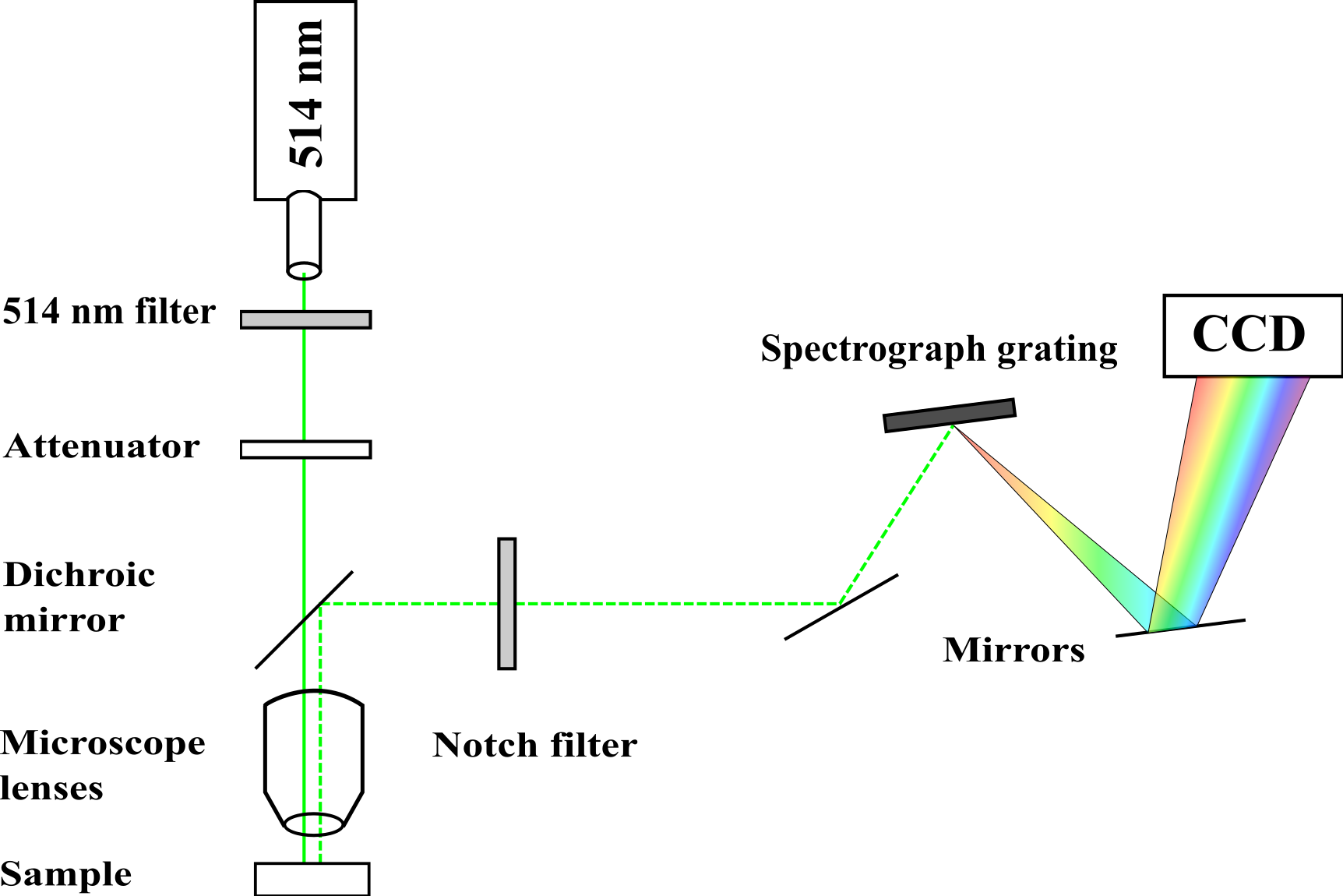}
    \caption{Microscope system for backscattering Raman spectroscopy.}
    \label{fig:RamanSystem}
\end{figure}

In this work, the labRAM HR system from \textit{Horiba} with a 514 nm laser is used to perform backscattering Raman spectroscopy. This system, illustrated in Fig. \ref{fig:RamanSystem}, is based on a confocal microscope with a filter and an attenuator on the incident light path while a spectrograph grating coupled with mirrors is used to discriminate spatially the energy components on a charge-coupled device (CCD) of the scattered light.

\begin{figure}[h!]
    \centering
    \includegraphics[width=1\linewidth]{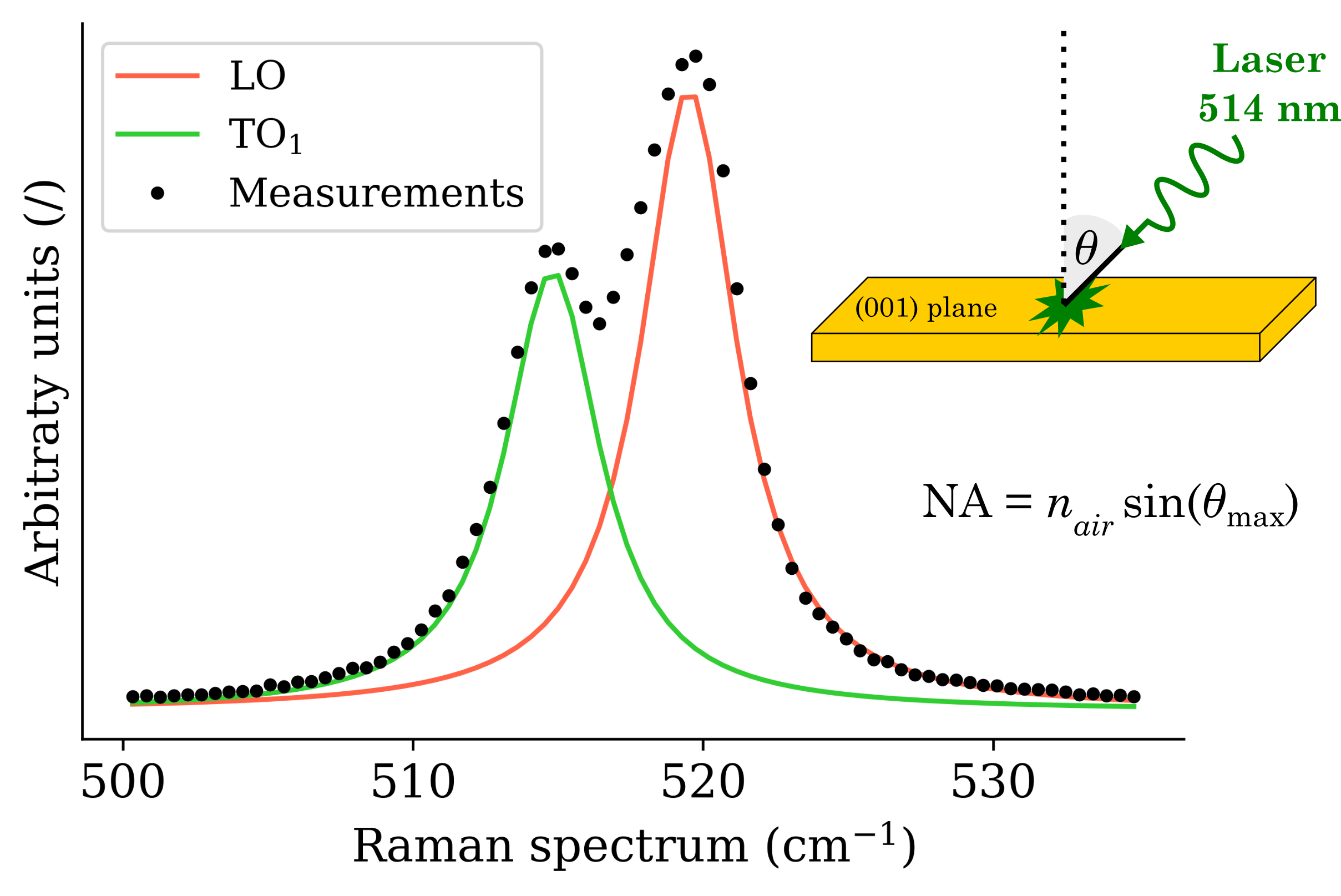}   
    \caption{Illustration of the Lorentzian fitting to retrieve the LO and TO$_1$ peaks in strained silicon samples. The transverse mode is observed thanks to the high numerical aperture (NA) of the objective that leads to a maximum incident angle ($\theta_{max}$) of 70° in the air ($n_{air}$=1).}
    \label{fig:LOTOpeaks}
\end{figure}

According to the selection rule, only the LO mode should be observed with the laser polarized in the [110] direction across the (001) crystal plane. However, the high numerical aperture (NA) of the microscope objective (MPLAPON100x from \textit{Olympus}) allows us to retrieve the TO$_1$ mode thanks to tilted components of the light beam hitting the material with a certain angle $\theta$ (cfr. Fig. \ref{fig:LOTOpeaks}). 
Indeed, input light with an incident angle ($\theta_{max}$) up to 70° can be collected by the objective with a numerical aperture of 0.95. The spectrum can then be fitted with Lorentzian functions to retrieve the positions of the Raman peaks \cite{Pelikan1994}.

\begin{figure}[h!]
    \centering
    \includegraphics[width=1\linewidth]{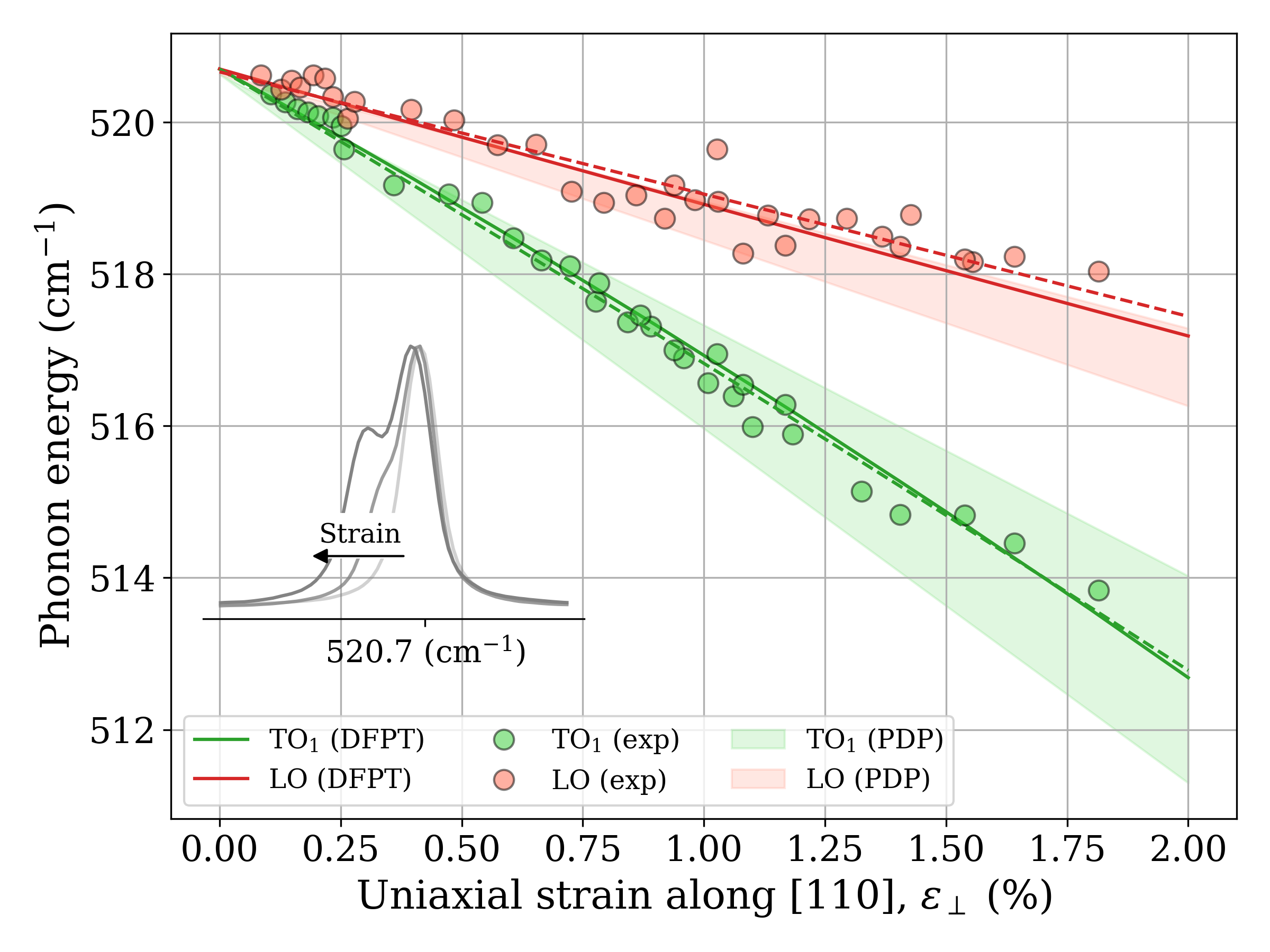}
    \caption{Experimental results of the shift of the LO (red dots) and TO$_1$ (green dots) Raman peaks of strained silicon. The measured data with fitting curves (dashed line) are compared with the DFPT simulation curves (continuous line) while the confidence intervals of PDP theory computed using the parameters from Table \ref{tab:pdpcoeff} are highlighted with filled areas. An illustration of the raw data is presented in the inset.}
    \label{fig:MainResults}
\end{figure}

The experimental results are displayed in Fig. \ref{fig:MainResults}. The results are compared with the DFPT simulation results while the confidence intervals due to the phonon deformation potentials found in the literature are highlighted. The strain-shift coefficients obtained experimentally are -160.99 cm$^{-1}$ for the LO mode and -414.97 cm$^{-1}$ for the TO$_1$ mode. The results for the TO peaks bring good agreement between the PDP theory, the DFPT simulations and the experimental work. The trend is however more spread for the LO mode but still presents an acceptable prediction of the Raman shift.



\section{Conclusion}
In this work, a first-principles method based on the density functional perturbation theory has been presented to compute the positions of the one-phonon Raman peaks in uniaxially strained silicon along the [110] crystal direction. The results have been discussed with regard to the literature using the linear model of the phonon deformation potentials theory and a weak non-linear behavior has been found. The simulated energy shifts of -175.77 cm$^{-1}$ for the LO mode, -400.85 cm$^{-1}$ for the TO$_1$ and 121.91 cm$^{-1}$ for the TO$_2$ one are in good agreement with other works. The experimental validation has been done on silicon microbeams fabricated using a lab-on-chip approach with silicon nitride actuators. These highly-strained beams (up to 2\%) were used to validate the first-principle approach for the prediction of the phonon energy shift in highly-strained material.
Scanning electron microscopy has been performed to retrieve the strain levels of the sample while the Raman peak positions of LO and TO$_1$ modes have been observed with a backscattering configuration under linearly polarized light. The experiment has shown strain-shift coefficients of -160.99 cm$^{-1}$ (LO) and -141.97 cm$^{-1}$ (TO$_1$), consistent with the DFPT results.

\section*{Aknowledgements}

Computational resources have been provided by the supercomputing facilities of the Université catholique de Louvain (CISM/UCL) and the Consortium des Équipements de Calcul Intensif en Fédération Wallonie Bruxelles (CÉCI) funded by the Fond de la Recherche Scientifique de Belgique (F.R.S.-FNRS) under convention 2.5020.11 and by the Walloon Region. M.-S. Colla acknowledges the financial support of National Fund for Scientific Research (FNRS), Belgium. N. Roisin acknowledges the help received for the first-principles simulations from G.-M. Rignanese and G. Brunin from Institute of Condensed Matter and Nanosciences (IMCN), Belgium.

\section*{Competing Interests}
The authors have no relevant financial or non-financial interests to disclose.

\section*{Data Availability}
All data generated or analysed during this study are included in this published article.

\section*{Author Contributions}

All authors contributed to the study conception and design. Material preparation, data collection and analysis were performed by Nicolas Roisin and Marie-Stéphane Colla. The first draft of the manuscript was written by Nicolas Roisin and Marie-Stéphane Colla and all authors commented on previous versions of the manuscript. All authors read and approved the final manuscript.

\bibliographystyle{elsarticle-num}
\bibliography{biblio.bib}

\end{document}